\documentclass{article}
\pdfoutput=1
\usepackage{PRIMEarxiv}
\usepackage{tensor}
\newcommand{\dV}{
    d^4 x
}
\newcommand{\half}{
    \frac{1}{2}
}
\newcommand{\beq}{\begin{equation}}
        \newcommand{\eeq}{\end{equation}}
\newcommand{\beqs}{\begin{eqnarray}}
        \newcommand{\eeqs}{\end{eqnarray}}

\newcommand{\pd}[1]{\tensor{\partial}{#1}}

\newcommand{\covard}[1]{
    \tensor{\nabla}{#1}
}

\newcommand{\g}[1]{
    \tensor{g}{ #1 }
}
\newcommand{\metric}[1]{
    \tensor{g}{ #1 }
}

\newcommand{\ricci}[1]{
    \tensor{\mathcal{R}}{ #1 }
}
\newcommand{\T}[1]{
    \tensor{T}{ #1 }
}
\newcommand{\R}{
    \mathcal{R}
}
\newcommand{\Lagr}{
    \mathcal{L}
}
\newcommand{\fR}{
    f(\mathcal{R})
}
\newcommand{\dfR}{
    f'(\mathcal{R})
}

\usepackage[utf8]{inputenc} 
\usepackage[T1]{fontenc}    
\usepackage{hyperref}       
\usepackage{url}            
\usepackage{booktabs}       
\usepackage{amsfonts}       
\usepackage{nicefrac}       
\usepackage{microtype}      
\usepackage{lipsum}         
\usepackage{fancyhdr}       
\usepackage{graphicx}       
\usepackage{doi}
\usepackage{amsmath}
\usepackage{amssymb}
\usepackage{subfig}
\usepackage{todonotes}
\graphicspath{{plots/}}     

\renewcommand{\eqref}[1]{Eq.~(\ref{#1})}
\newcommand{\figref}[1]{Fig.~(\ref{#1})}

\title{Neutral particle motion around a Schwarzschild-de Sitter Black Hole in $\fR$ gravity
}

\author{
  Devansh~Shukla \\
  Sardar Vallabhbhai National Institute of Technology, Surat, India \\
  \texttt{devanshshukla99@gmail.com} \\
  \And
  Kamlesh~Pathak \\
  Sardar Vallabhbhai National Institute of Technology, Surat, India \\
  \texttt{knp@phy.svnit.ac.in}
}

\begin{document}
\maketitle
\begin{abstract}
  This article investigates the presence of a static spherically symmetric solution in the metric f(R) gravity.
  Consequently, we have examined the presence of horizons for the extreme and hyperextreme Schwarzschild-de Sitter solution.
  Further, we have investigated the orbital motion of a time-like particle around the Schwarzschild-dS solution by forming the constraints for the existence of circular orbits and have subsequently developed an approximation to the innermost stable circular orbit (ISCO).
\end{abstract}


\section{Introduction}
\paragraph{}
Einstein initially proposed the theory of General Relativity in 1915\cite{Einstein:1915ca,Einstein:1916vd}. Now, over a century later, concerns regarding its limitations are increasingly prominent.
Although, Einstein's General Relativity has stood observational and experimental scrutiny, it falls short of explaining recent cosmological and astrophysical data such as galactic rotation curves and cosmic inflation.

Recent cosmological evidence, primarily derived from type Ia supernovae explosions, indicates that the universe is indeed undergoing acceleration\cite{Riess_1998}. However, GR fails to explain this without invoking minuscule cosmological constant or a form of dark energy\cite{Charmousis2009, RevModPhys.82.451}.
The most basic model that which fits the observational data is the  concordance or standard $\Lambda$ Cold Dark Matter($\Lambda$CDM) model with some inflationary scenario, with a field called inflaton\cite{RevModPhys.82.451}.
Even so, it is also plagued by various challenges. It neither explains the nature of the inflationary field nor the nature of dark energy or matter.
Furthermore, it is plagued by the coincidence problem, horizon problem, and some physicists find the degree of fine-tuning to be disconcerting\cite{carroll1997lecture, PERIVOLAROPOULOS2022101659}.

Additionally, it is expected that Einstein's theory breaks down at extremely high energy levels near the Planck's scale, where higher order curvature terms dominates and can no longer be neglected\cite{Charmousis2009}.
Thus, providing sufficient motivation to study modified gravity theories which can consider those higher order curvature terms, like f(R) theories of gravity.

$\fR$ theories of gravity is a family of theories, each defined with a different function of the Ricci scalar, $\R$.
These theories arise by a straightforward generalization of the Lagrangian in the Einstein-Hilbert action to become a general function\cite{RevModPhys.82.451} of $\R$ as:
\begin{equation}
  S = \dfrac{1}{2\kappa} \int d^4x \sqrt{-g} \fR
\end{equation}
It was first propsoed by Hans Adolf Buchdahl in 1970s and later matured and used by Starobinsky for explaininig the cosmic inflation\cite{STAROBINSKY198099}.
$\fR$ theories of gravity are an interesting and relatively simple modification to Einstein's General Relativity\cite{Einstein:1915ca, Einstein:1916vd}.
Although, it is relatively straightforward to handle, its action is adequately comprehensive to encompass some fundamental attributes of higher-order gravity.

There exist three versions of f(R) gravity. The first version is defined using the metric formalism\cite{PhysRevD.74.086005}, the second version is known as Palatini's f(R) gravity\cite{Capozziello_2007} and is defined in the Palatini formalism, and the third version is the metric-affine f(R) gravity\cite{SOTIRIOU2007935}.
The metric-affine f(R) gravity represents the most general and comprehensive form, which can be simplified to metric or Palatini f(R) gravity given specific assumptions.

In this article, we will utilize the metric formalism for f(R) gravity.
However, as obvious from previous literatures\cite{PhysRevD.80.124011}, the equations prove to be a significant challenge for analytical solutions without additional simplifying assumptions.
So, we will employ the popular option of using constant scalar curvature($R_0$) with \eqref{eq:fr_conditions}.
\begin{equation}
  f(R_0) = \pd{_r} f(R_0) = 0  \label{eq:fr_conditions}
\end{equation}
Further, we will assume a static spherically symmetric metric to obtain the line-element for a Schwarzschild-de Sitter black hole in f(R) gravity.
While the general relativity does not always allow for the consideration of circular motion of a test particle, it remains valuable in numerous theoretical investigations such as the study of spacetime geometry, understanding accretion disks, and analyzing geodesic motion.
Therefore, we will outline the constraints and solutions pertaining to the horizons of the Schwarzschild-dS black hole\cite{griffiths_podolský_2009}, as well as the existence of circular orbits such as the marginally stable circular orbit (MSCO) and the inner-most stable circular orbit (ISCO)\cite{1974ApJ191499P, wald2010general}.

The paper is organized as follows. In section 2, we provide a brief explanation of the metric formulism of f(R) gravity and develop the field equations. In section 3, we analyze the ansatz model for the static spherically symmetric solution and examine the metric.
Section 4 deals with computing the horizons for near-extreme Schwarzschild-de Sitter blackhole and existence of holes in hyperextreme case. In Section 5, we utilize the line-element obtained in Section 3 to develop an effective potential and develop the conditions for the existence of circular orbits. Later in the section, we develop the marginally stable circular orbits and prove it as the approximation to the innermost stable circular orbits for time like particles in f(R) gravity.
Finally, in section 6, we summarize the results.

We use the natural units $c=G=k_B=\hbar=1$ with the metric convention of $(-, +, +, +)$ thoughout the paper.

\section{The ansatz model}
\paragraph*{}
In order to obtain the ansatz model for $\fR$ gravity, we attempt to generalize the Einstein-Hilbert action by replacing the ricci scalar $\mathcal{R}$ by an arbitary function of the scalar curvature $\fR$.
The generalization is real and important and we use it to compute the Einstein-Hilbert action:

The modified Einstein-Hilbert action reads
\begin{equation}
  A_{g} = \frac{1}{2\kappa} \int \dV \sqrt{-g} \fR
\end{equation}
The total action reads:
\begin{equation}
  A = A_{g} + A_{m}
\end{equation}
where $A_{m}$ represents the matter action. We develop the field equation by varying the action with respect to the metric:
\begin{equation*}
  \dfrac{\delta A}{\delta \g{^\mu ^\nu}} = \frac{1}{2\kappa} \int \dfrac{\delta}{\delta \g{^\mu ^\nu}} \left\{ \dV \sqrt{-g} \fR \right\} + \dfrac{\delta A_m}{\delta \g{^\mu ^\nu}}
\end{equation*}
\begin{equation}
  \dfrac{\delta A_g}{\delta \g{^\mu ^\nu}} = \frac{1}{2\kappa} \int \dV \left\{ \left(\dfrac{\delta}{\delta \g{^\mu ^\nu}} \sqrt{-g}\right) \fR + \sqrt{-g} f'(\R) \dfrac{\delta \R}{\delta \g{^\mu ^\nu}} \right\}  \label{eq:gravity_action_variation}
\end{equation}
where $f'(\R) = \pd{_\R} \fR$

The variation of the scalar tensor is computed from the defination:
\begin{equation*}
  \dfrac{\delta \R}{\delta \g{^\mu ^\nu}} = \dfrac{\delta}{\delta \g{^\mu ^\nu}} \left(\g{^\mu ^\nu} \ricci{_\mu _\nu}\right)  
\end{equation*}
\begin{equation}
  \dfrac{\delta \R}{\delta \g{^\mu ^\nu}} = \ricci{_\mu _\nu} + \g{_\mu _\nu} \Box - \covard{_\mu} \covard{_\nu} \label{eq:riemannian_tensor_variation} 
\end{equation}
where $\Box \equiv \g{^\alpha ^\beta} \covard{_\alpha} \covard{_\beta} $

Using \eqref{eq:riemannian_tensor_variation} with \eqref{eq:gravity_action_variation}, we obtain:
\begin{equation}
  \delta A_{g} = \frac{1}{2\kappa} \int \dV \left\{ -\half \sqrt{-g} \g{_\mu _\nu} \delta \g{^\mu ^\nu} \fR + \sqrt{-g} \dfR \left\{\delta \g{^\mu ^\nu} \ricci{_\mu _\nu} + \g{_\mu _\nu} \Box \delta \g{^\mu ^\nu} - \covard{_\mu} \covard{_\nu} \delta\g{^\mu ^\nu}\right\} \right\}
\end{equation}
\begin{equation}
  \delta A_{g} = \frac{1}{2\kappa} \int \dV \sqrt{-g} \delta \g{^\mu ^\nu} \left\{\ricci{_\mu _\nu} \dfR - \half \g{_\mu _\nu} \fR - \left(\covard{_\mu} \covard{_\nu} - \g{_\mu _\nu} \Box\right)\dfR \right\}  \label{eq:sim_lagrangian_gravity}
\end{equation}
Using \eqref{eq:gravity_action_variation} and \eqref{eq:sim_lagrangian_gravity}, the field equations reads:
\begin{equation}
  \ricci{_\mu _\nu} \dfR - \half \g{_\mu _\nu} \fR - \left(\covard{_\mu} \covard{_\nu} - \g{_\mu _\nu} \Box\right)\dfR = \kappa \T{_\mu _\nu}
\end{equation}
with $\T{_\mu _\nu} =  \dfrac{-2}{\sqrt{-g}} \dfrac{\delta A_m}{\delta \g{^\mu ^\nu}}  \label{eq:stress_energy_tensor}$

\section{Static spherically symmetric solution}
In order ot obtain the Schwarzschild's line element, we assume an ansatz solution for the spherically symmetric solution characterised by the radial coordinate with the analomous red shift $\phi(r)$\cite{Calz_2018}:
\begin{equation}
  ds^2 = - e^{-2 \phi(r)} Z(r) dt^2 + Z^{-1}(r) dr^2 + r^2 d\Omega^2_k  \label{eq:sss_metric}
\end{equation}
Here $d\Omega^2_{k}$ is the metric of a constant curvature two-dimensional space, with three different possible topologies namely spherical ($k = 1$), flat ($k = 0$) and hyperbolic ($k = -1$).
\begin{equation}
  d\Omega_k^2 = \dfrac{1}{1-k\rho^2} d\rho^2 + \rho^2 d\phi^2
\end{equation}
Evaluating the Ricci scalar using the metric defined in \eqref{eq:sss_metric}:
\begin{equation}
  \mathcal{R} = \frac{2 k+r^2 \left(-Z''(r)\right)+Z(r) \left(2 r^2 \phi ''(r)-2 r^2 \phi '(r)^2+4 r \phi '(r)-2\right)+r Z'(r) \left(3 r \phi '(r)-4\right)}{r^2}
\end{equation}
where $Z'(r) = \partial_{r} Z(r)$ and $\phi'(r) = \partial_{r} \phi(r)$

Adopting a constant scalar curvature $\mathcal{R} = R_0$ and the simplest case of anamolous red-shift where $\phi(r)=\phi$,  this implies that $\phi'(r)$ and all further derivatives vanishes, leaving a trivial second-order differential equation \eqref{eq:sss_ricci_scalar_nophi}:
\begin{equation}
  R_0 = \frac{2 k - r^2 Z''(r)-4 r Z'(r)-2 Z(r)}{r^2}  \label{eq:sss_ricci_scalar_nophi}
\end{equation}
The solution of \eqref{eq:sss_ricci_scalar_nophi} can be expressed as \eqref{eq:sss_ricci_scalar_solution}.
It is worth noting that the obtained solution bears resemblance to the Nariai solution, which is discussed further in the next section.
\begin{equation}
  Z(r) = k + \dfrac{C_1}{r} + \dfrac{C_2}{r^2} - \frac{r^2 R_0}{12}  \label{eq:sss_ricci_scalar_solution}
\end{equation}
Here, anti-de Sitter (AdS) and de Sitter (dS) space are denoted by the appropriate values of $k$ and the scalar curvature $R_0$ respectively.
This article exclusively focuses on the de Sitter treatment, hence, $R_0>0$ and $k=1$.

The solution for our static spherically symmetric line element with reads:
\begin{equation}
  ds^2 = -\left(1 + \dfrac{C_1}{r} + \dfrac{C_2}{r^2} - \dfrac{r^2 R_0}{12}\right) dt^2 + \left(1 + \dfrac{C_1}{r} + \dfrac{C_2}{r^2} - \dfrac{r^2 R_0}{12}\right)^{-1} dr^2 + r^2 d\Omega^2 \label{eq:sss_metric_horizon}
\end{equation}
In the upcoming section, we will calculate the value of horizons using the computed line element.

\section{Investigation of horizons}
By utilizing the static spherically symmetric line element derived in the previous section, we can express $\metric{_t_t}$ as:
\begin{equation}
    \metric{_t_t}(r) = - \left(1 + \dfrac{C_1}{r} + \dfrac{C_2}{r^2} - \dfrac{r^2 R_0}{12}\right) \label{eq:gtt}
\end{equation}
The \eqref{eq:gtt} bears resemblance to the Schwarzschild-dS solution in vanilla general relativity.

The Schwarzschild-de Sitter is the generalization of the Schwarzschild solution with mass parameter $m$ and an arbitary cosmological constant $\Lambda$\cite{Podolsky_1999, griffiths_podolský_2009}. The metric for this case was discovered by Kottler\cite{kottler1918}, Weyl\cite{weyl1919b} and Trefftz\cite{Trefftz1922} and can be written as:
\begin{equation}
    ds^2 = - \left(1 - \dfrac{2m}{r} - \dfrac{\Lambda}{3} r^2\right) dt^2 + \left(1 - \dfrac{2m}{r} - \dfrac{\Lambda}{3} r^2\right)^{-1} dr^2 + r^2 \left(d\theta^2 + \sin^2\theta d\phi^2\right) \label{eq:schwarzschild_desitter}
\end{equation}
This solution reduces to Schwarzschild metric when $\Lambda=0$ and to de Sitter or anti-de Sitter metric in their spherically symmetric forms when $m=0$\cite{griffiths_podolský_2009}.

By comparing the $\metric{_t_t}$ component in our SSS line-element with that of the Schwarzschild-dS solution, we can assume that $C_2$ vanishes.
Although, eliminating $C_2$ will reduce the number of independent variables, but it will significantly simplify our calculations from a quartic equation to a cubic equation.
Furthermore, it is demostrated in multiple articles that $C_2=Q$ and corresponds to Reissner-Nordstr\"{o}m solution as charge $Q$\cite{PhysRevD.80.124011, S0218271816500784}.

Additonally, given that the theory must conform to standard general relativity when there is no curvature $(R_0)$, we can infer that $C_1 = -2m$ and $\dfrac{R_0}{4} = \Lambda$ \cite{PhysRevD.80.124011}

The $\metric{_t _t}(r)$ becomes:
\begin{equation}
    \metric{_t_t}(r) = \dfrac{1}{12r}\left(R_0 r^3 - 12r + 24m\right) \label{eq:gtt_1}
\end{equation}
Our primary focus is on understanding the characteristics of $\metric{_t _t}$. Therefore, we will exclusively concentrate on the cubic polynomial at this moment:
\begin{equation}
    R_0 r^3 - 12r + 24m \label{eq:gtt_cubic_polynomial}
\end{equation}

The discriminant for the cubic polynomial in \eqref{eq:gtt_cubic_polynomial} can be expressed as:
\begin{equation}
    \Delta_3 = 432 R_0 \left(16 - 9 (4 m^2) R_0\right)
\end{equation}
Trivially, it can be noted that for $\Delta_3 > 0$: there are three distinct real roots, whereas for $\Delta_3<0$: there is one real root and two complex conjuagte roots and for $\Delta_3=0$: all roots are real but two are repeated roots.

In order to achieve a positive $\Delta_3$, the following condition must be true:
\begin{equation}
    R_0 \left(4 - 9 m^2 R_0\right) > 0
\end{equation}

Therefore, for a deSitter spacetime, the relation between $C_1$ and $R_0$ to obtain a postitive discriminant is given as:
\begin{equation}
    \boxed{0 < \dfrac{9 m^2}{4} R_0 < 1} \label{eq:constraint}
\end{equation}

Following the constraint in the above equation, we can compute the roots for the \eqref{eq:gtt_cubic_polynomial} using the trigonometric method for solving depressed cubic equations, as originally derived by François Viète\cite{nickalls_2006}.
\begin{eqnarray}
    \begin{aligned}
        r_1 & = 4 \sqrt{\frac{1}{R_0}} \cos \left(\frac{1}{3} \arccos\left(-\frac{3}{2} m \sqrt{R_0}\right)\right)                 \\
        r_2 & = -4 \sqrt{\frac{1}{R_0}} \sin \left(\frac{\pi }{6}-\frac{1}{3} \arccos\left(-\frac{3}{2} m \sqrt{R_0}\right)\right) \\
        r_3 & = -4 \sqrt{\frac{1}{R_0}} \sin \left(\frac{1}{3} \arccos\left(-\frac{3}{2} m \sqrt{R_0}\right)+\frac{\pi }{6}\right)
    \end{aligned} \label{eq:_new_gtt_roots}
\end{eqnarray}

Furthermore, since \eqref{eq:gtt_cubic_polynomial} is a depressed cubic polynomial with no $2^{st}$ order term, the sum of its roots must be zero. Consequently, the third root will be negative and equal to the sum of the other two roots.
\begin{figure}[!h]
    \centering
    \includegraphics[scale=0.5]{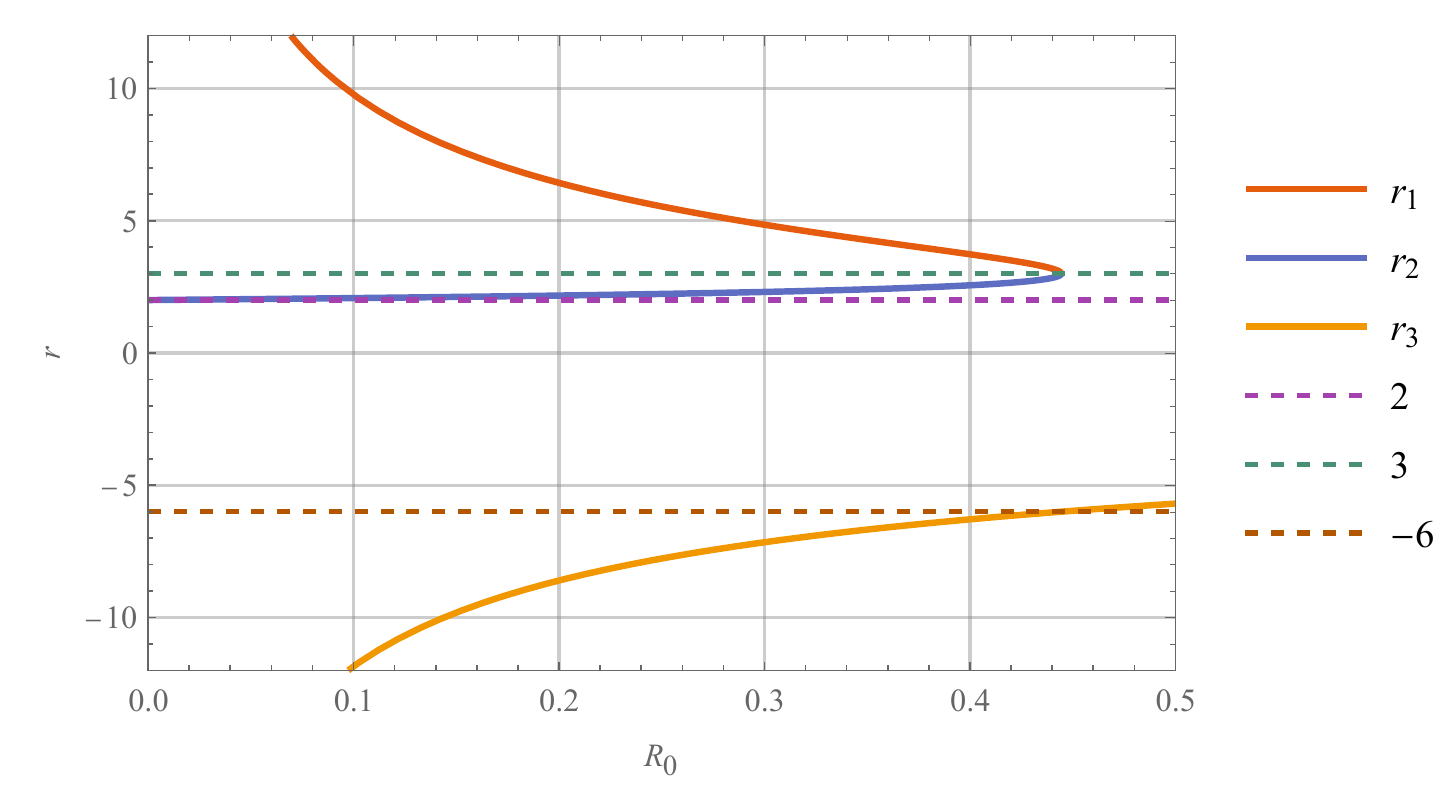}
    \caption{Plot for $r$ vs $R_0$ assuming unit mass(m=1) and deSitter space-time} \label{fig:r_vs_R0}
\end{figure}

In \figref{fig:r_vs_R0}, we are assuming the value of mass to be unity for simplicity. However, this assumption does not result in any loss of degrees of freedom due to the constraint described in \eqref{eq:constraint}.

Upon analyzing \figref{fig:r_vs_R0}, it is evident that initially the value of $r_1$ is infinitely large while the value of $r_2$ is $2$.
As both approach the extremum limit in deSitter space-time, they both collapse to a shared value of $3$.
This is consistant with the concept that $r_1$ represents the cosmological horizon and $r_2$ represents the Schwarzschild black hole horizon.

There are no real horizons distances in the hyperextreme case $\left(\dfrac{9R_0m^2}{4} > 1\right)$ so there are no blackholes just naked singularity without horizons\cite{Geyer1980, griffiths_podolský_2009}.

\section{Circular orbit solution}
We will now work towards computing the innermost stable circular orbit for our static spherically symmetric solution.
\begin{equation}
    Z(r) = 1 + \dfrac{C_1}{r} + \dfrac{C_2}{r^2} - \dfrac{r^2 R_0}{12}
\end{equation}
Let us assume a test particle of a non-negative $\mu$ in the equitorial frame($\theta=\pi/2$). The line element reads:
\begin{equation}
    ds^2 = \g{_\alpha _\beta} dx^\alpha dx^\beta = -Z(r) dt^2 + \dfrac{1}{Z(r)} dr^2 + r^2 d\phi^2 = \epsilon \label{eq:spacetime_interval}
\end{equation}
where $\epsilon=-1, 0, 1$ for timelike, null-like and spacelike trajectories respectively.

The lagrangian of a free particle in a potential-less spacetime reads:
\begin{equation}
    \Lagr = \mu \sqrt{\g{_\alpha _\beta} \dot{x}^\alpha \dot{x}^\beta}
\end{equation}
Considering our lagrangian is independent of $t$ and $\phi$, the first integrals of motion obtained using the Euler-Lagrange equation reads:
\begin{equation}
    \mu Z(r) \dot{t} = E; \quad \mu r^2 \dot{\phi} = L \label{eq:first_intergrals_lagr}
\end{equation}
where $E$ and $L$ represent the energy and angular momentum of the test particle respectively.

Substituting \eqref{eq:first_intergrals_lagr} into \eqref{eq:spacetime_interval}:
\begin{equation}
    -\dfrac{E^2}{\mu^2 Z(r)} + \dfrac{\dot{r}^2}{Z(r)} + \dfrac{L^2}{\mu^2 r^2} = \epsilon
\end{equation}
\begin{equation}
    \dot{r}^2 + V_{\text{eff}}^2 = \dfrac{E^2}{\mu^2}
\end{equation}
with the effective potential represented by:
\begin{equation}
    V_{\text{eff}} = \sqrt{\left(\dfrac{L^2}{\mu^2 r^2} - \epsilon\right)Z(r)}\label{eq:effective_potential}
\end{equation}
The circular orbits are determined by the condition that $\dot{r}=0$. These circular orbits sit at critical points of the effective potential $V_{\text{eff}}$, or simply $V_{\text{eff}}'(r_o)=0$\cite{Howes_1979, Berenstein_2021}.
The extremum values of $V_{\text{eff}}'$ determines both the stable and unstable circular orbits for the test particle. For the orbit to achieve classical stability, it is necesaary that $r_o$ statifies $V_{\text{eff}}''(r_0)>0$\cite{Berenstein_2021, Hussain_2016}.

The notion of innermost stable circular orbits (ISCO) $r_{\text{isco}}$ corresponds to the marginally stable circular orbits\cite{Howes_1979, nasereldin2019boundary} which satisfies both: $V_\text{eff}'(r_\text{isco}) = V_{\text{eff}}''(r_{\text{isco}})=0$

Solving for $\pd{_r} V_{\text{eff}} = 0$, we obtain the value of the angular momentum and energy as:
\begin{equation}
    \dfrac{L^2}{r^2 \mu^2} = \frac{\epsilon  \left(6 C_1 r+12 C_2+r^4 R_0\right)}{6 \left(3 C_1 r+4 C_2+2 r^2\right)}
\end{equation}
\begin{equation}
    \dfrac{E^2}{\mu^2} = -\frac{\epsilon  \left(12 C_1 r+12 C_2-r^4 R_0+12 r^2\right){}^2}{72 r^2 \left(3 C_1 r+4 C_2+2 r^2\right)}
\end{equation}

By choosing $\epsilon=-1$, we restrict our analysis to exclusively timelike orbits. To streamline the calculation process, we will employ the assumption derived from the previous section: $C_1=-2m$ and $C_2=0$

In this case, the corresponding angular momentum and potential can be expressed as:
\begin{equation}
    \dfrac{L^2}{\mu^2} = \dfrac{r^2}{12} \left(\dfrac{r^3 R_0-12 m}{3 m-r} \right) \label{eq:timelike_angular_momentum}
\end{equation}
\begin{equation}
    \dfrac{E^2}{\mu^2} = -\frac{\left(24 m+r^3 R_0-12 r\right){}^2}{144 r (3 m-r)} \label{eq:timelike_energy}
\end{equation}
Analyzing \eqref{eq:timelike_angular_momentum} indicates that in order for $L^2/\mu^2$ to have a positive value when $r$ exceeds $3m$, the expression $r^3 R_0 - 12m$ must be negative.
Therefore, we obtain the following constraint for the radial parameter of a circular orbit:
\begin{equation}
    3m<r<\left(\dfrac{12m}{R_0}\right)^{1/3}
\end{equation}
The inverse scenario when $r<3m$, the expression $(r^3 R_0 - 12 m)$ must be positive.
By substituting the max value of scalar curvature permitted by constraint \eqref{eq:constraint}, it becomes evident that this scenario is not possible.

Now, in order to compute the innermost stable circular orbit we again extremise the potential $V_{\text{eff}}$, given by \eqref{eq:effective_potential}, and solve for $r$:
\begin{equation}
    \left(r^3 R_0-12 r+24 m\right) \left(r^3 R_0 (4 r-15 m)+12 m (6 m-r)\right)=0 \label{eq:_isco_eq_1}
\end{equation}

The first part of \eqref{eq:_isco_eq_1} is simply the polynomial we solved in the earlier section and obtained the blackhole and cosmological horizons value.
In this section, we will focus on the second part of the equation, the fourth order polynomial for finding the circular orbits:
\begin{equation}
    \left(r^3 R_0 (4 r-15 m)+12 m (6 m-r)\right)=0 \label{eq:_isco_eq_2}
\end{equation}

In order to ascertain the nature of roots, we first compute the discriminant for the fourth order polynomial \eqref{eq:_isco_eq_2} as:
\begin{equation}
    \Delta_4 = -8957952 m^4 R_0^2 \left(9 m^2 R_0-4\right) \left(5625 m^2 R_0-16\right)
\end{equation}

Trivially, it can be noted that for: $\Delta_4>0$: there exists four real roots or 2 pairs of conjugated complex roots, $\Delta_4=0$: there exists a minimum 2 complex/real roots are repeated, $\Delta_4<0$: there exists a pair of conjugated complex roots and two real roots, all of these are distinct.

So, in hopes of obtaining real and distinct roots, assuming $m>0$ and $R_0>0$, we see that the discriminant is negative for $\left(9 m^2 R_0-4\right)<0$ and $\left(5625 m^2 R_0-16\right)<0$, or simply:
\begin{equation}
    0 < m^2 R_0 < \dfrac{16}{5625} \label{eq:isco_limiting}
\end{equation}

Now, by solving the quartic equation given in \eqref{eq:_isco_eq_2} using Mathematica\cite{Mathematica}, we find the real roots as:
\begin{eqnarray}
    \begin{aligned}
        r_{\text{o}_1} & = \dfrac{15 m}{16} + \dfrac{1}{16}\left[{9 m^2 \left(96 \;{6}^{1/3} \; {\Xi}^{-1} + 25\right)+\frac{16\ 6^{2/3}\; {\Xi}}{R_0}}\right]^{1/2}                                                                                                                                 \\
                       & - \dfrac{1}{16}\left[{450 m^2-\frac{16\ 6^{2/3} \;{\Xi}}{R_0}} - 864 \;{6}^{1/3}\; m^2 \; {\Xi}^{-1} + \frac{6 m \left(1125 m^2 R_0+512\right)}{R_0 \left[{9 m^2 \left(96\; {6}^{1/3}\; {\Xi}^{-1} + 25\right)+\frac{16\; 6^{2/3}\; {\Xi}}{R_0}}\right]^{1/2}}\right]^{1/2} \\
        r_{\text{o}_2} & = \dfrac{15 m}{16} + \dfrac{1}{16}\left[{9 m^2 \left(96 \;{6}^{1/3} \; {\Xi}^{-1} + 25\right)+\frac{16\ 6^{2/3}\; {\Xi}}{R_0}}\right]^{1/2}                                                                                                                                 \\
                       & + \dfrac{1}{16}\left[{450 m^2-\frac{16\ 6^{2/3} \;{\Xi}}{R_0}} - 864 \;{6}^{1/3}\; m^2 \; {\Xi}^{-1} + \frac{6 m \left(1125 m^2 R_0+512\right)}{R_0 \left[{9 m^2 \left(96\; {6}^{1/3}\; {\Xi}^{-1} + 25\right)+\frac{16\; 6^{2/3}\; {\Xi}}{R_0}}\right]^{1/2}}\right]^{1/2}
    \end{aligned}\label{eq:_isco_eq_2_roots}
\end{eqnarray}
where $\Xi = \left[m^2 R_0 \left(225 m^2 R_0+\sqrt{50625 m^4 R_0^2-22644 m^2 R_0+64}+8\right)\right]^{1/3}$

\begin{figure}[!h]
    \centering
    \includegraphics[scale=0.65]{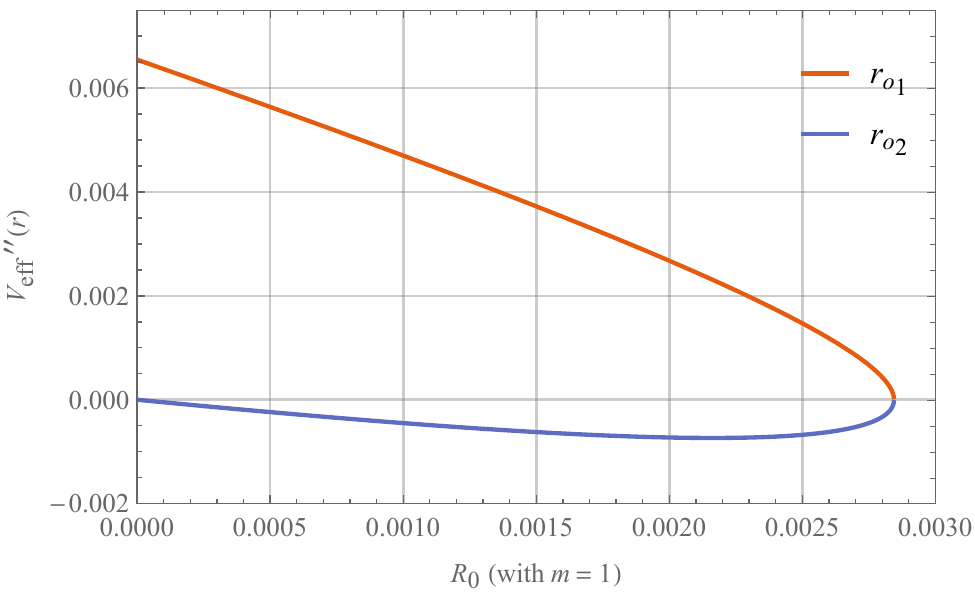}
    \caption{Plot for $V_{\text{eff}}''$ vs $R_0$ for both roots $r_{o_1}$ and $r_{o_2}$} \label{fig:ddenergyR0}
\end{figure}

The second derivative of the effective potential, $V_{\text{eff}}''(r)$ can be represented as \eqref{eq:ddveff}.
In order for the circular orbit solutions to be stable, it is necessary that the second derivative of the effective potential must be positive, $V_{\text{eff}}''>0$.
However, the analytical verification for $V_{\text{eff}''(r_{\text{msco}})}>0$ is difficult; instead, we can easily demonstrate it through \figref{fig:ddenergyR0}
\begin{equation}
    V_{\text{eff}}''(r) = \dfrac{\left(24 m+r^3 R_0-12 r\right) \left(12 m \left(54 m^2-39 m r+4 r^2\right)+r^3 R_0 \left(135 m^2-60 m r+8 r^2\right)\right)}{48 (r (r-3 m))^{5/2} \;\left| R_0 r^3-12 r+24 m\right| } \label{eq:ddveff}
\end{equation}

As evident from \figref{fig:ddenergyR0}, $r_{o_1}$ represents a stable circular orbit whereas $r_{o_2}$ is unstable and would inevitably plunge into the black hole.

Therefore, the marginally stable circular orbit is:
\begin{eqnarray}
    \begin{aligned}
        r_{\text{msco}} & = \dfrac{15 m}{16} + \dfrac{1}{16}\left[{9 m^2 \left(96 \;{6}^{1/3} \; {\Xi}^{-1} + 25\right)+\frac{16\ 6^{2/3}\; {\Xi}}{R_0}}\right]^{1/2}                                                                                                                                 \\
                        & - \dfrac{1}{16}\left[{450 m^2-\frac{16\ 6^{2/3} \;{\Xi}}{R_0}} - 864 \;{6}^{1/3}\; m^2 \; {\Xi}^{-1} + \frac{6 m \left(1125 m^2 R_0+512\right)}{R_0 \left[{9 m^2 \left(96\; {6}^{1/3}\; {\Xi}^{-1} + 25\right)+\frac{16\; 6^{2/3}\; {\Xi}}{R_0}}\right]^{1/2}}\right]^{1/2} \\
    \end{aligned}
\end{eqnarray}

Furthermore, under $R_0\rightarrow0$, the modified theory has to converge back to standard general relativity. Applying this limit to $r_\text{msco}$ with $m=1$ yields a value of $6.00016$, which is an extremely close approximation to Einstein's relativty ISCO of $6.00$.
Additonally, as shown in \figref{fig:ddenergyR0}, the $V_\text{eff}''$ does not vanish but is a good approximation to zero.
As a result, $r_\text{msco}$, the solution we calculated, is a reasonable approximation to $r_\text{isco}$.

\begin{eqnarray}
    \begin{aligned}
        r_{\text{isco}} & \approx \dfrac{15 m}{16} + \dfrac{1}{16}\left[{9 m^2 \left(96 \;{6}^{1/3} \; {\Xi}^{-1} + 25\right)+\frac{16\ 6^{2/3}\; {\Xi}}{R_0}}\right]^{1/2}                                                                                                                           \\
                        & - \dfrac{1}{16}\left[{450 m^2-\frac{16\ 6^{2/3} \;{\Xi}}{R_0}} - 864 \;{6}^{1/3}\; m^2 \; {\Xi}^{-1} + \frac{6 m \left(1125 m^2 R_0+512\right)}{R_0 \left[{9 m^2 \left(96\; {6}^{1/3}\; {\Xi}^{-1} + 25\right)+\frac{16\; 6^{2/3}\; {\Xi}}{R_0}}\right]^{1/2}}\right]^{1/2} \\
    \end{aligned}
\end{eqnarray}

\begin{figure}[!h]
    \centering
    \subfloat[$r_{\text{isco}}$ as a function of $R_0$ with constant $m$]{\includegraphics[width=8.2cm]{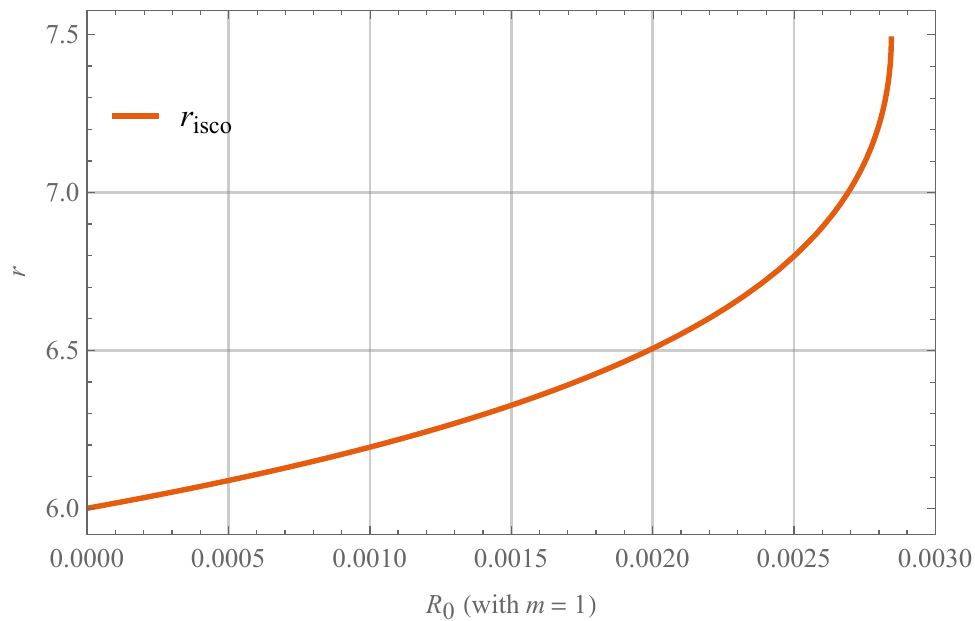}\label{fig:riscoR0}}
    \subfloat[$r_{\text{isco}}$ as a function of $m$ with constant $R_0$]{\includegraphics[width=8cm]{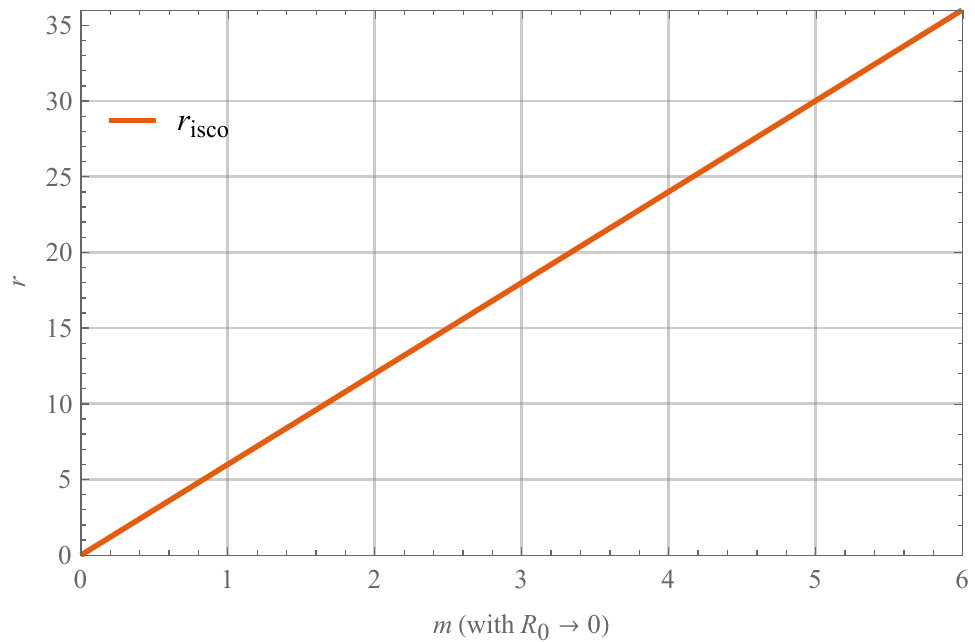}\label{fig:riscom}}
    \caption{Approximated ISCO solution as a function of $R_0$ with constant m and vice-versa}\label{fig:risco}
\end{figure}

\figref{fig:riscoR0} displays the graph of our approximated solution for the Innermost Stable Circular Orbit (ISCO) as a function of $R_0$ whereas \figref{fig:riscom} displays ISCO as a funciton of $m$ considering $R_0\rightarrow0$, assuming unit mass (m = 1).

It is evident from \figref{fig:risco} that the approximated solution for ISCO behaves quadratically with $R_0$ for $m\rightarrow1$ and linearly with $m$ for $R_0\rightarrow0$.
It is important to remember that the limits determined by \eqref{eq:isco_limiting} constrain the ISCO solution.

\begin{figure}[!ht]
    \centering
    \subfloat[Angular momentum per unit mass squared vs r]{\includegraphics[scale=0.45]{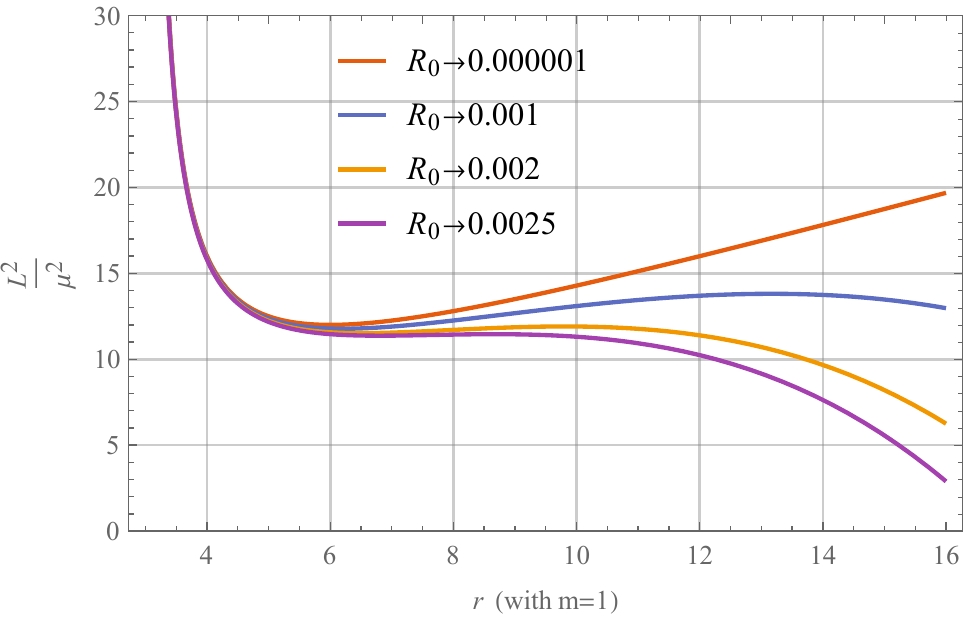}}
    \subfloat[Effective potential vs r]{\includegraphics[scale=0.44]{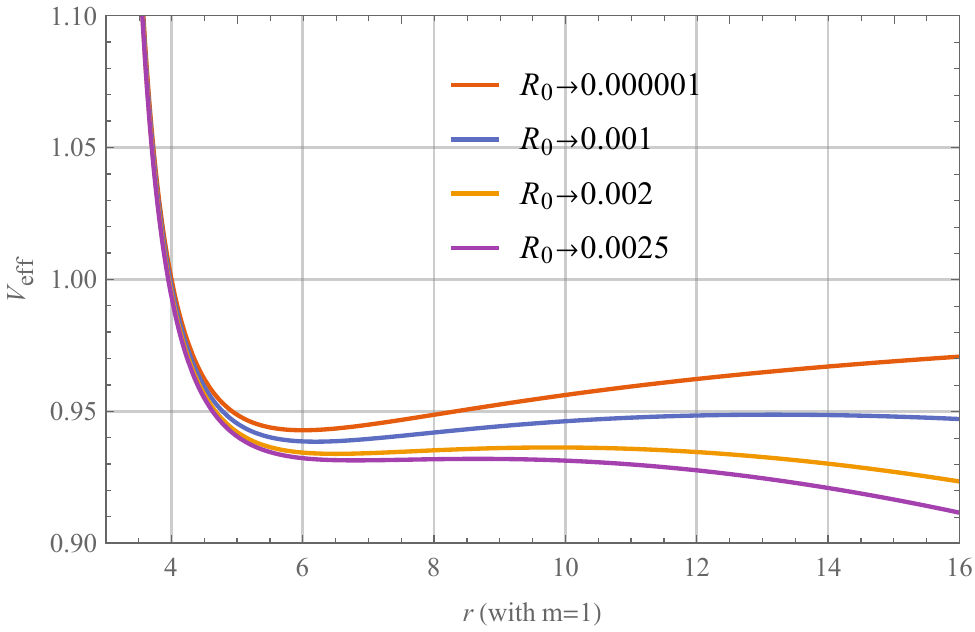}} \\
    \subfloat[Derivative of effective potential vs r]{\includegraphics[scale=0.44]{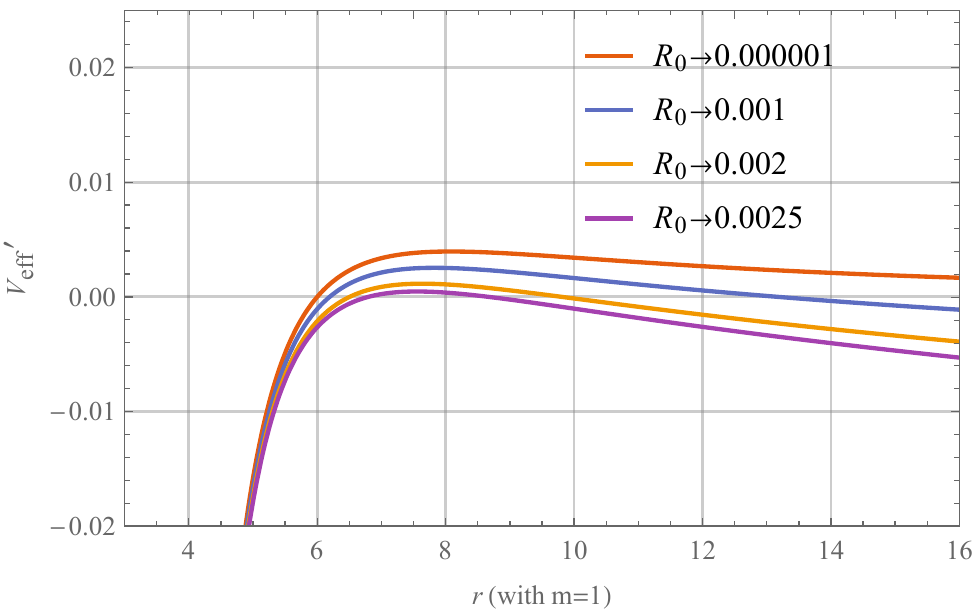}}
    \subfloat[Double derivative of effective potential vs r]{\includegraphics[scale=0.44]{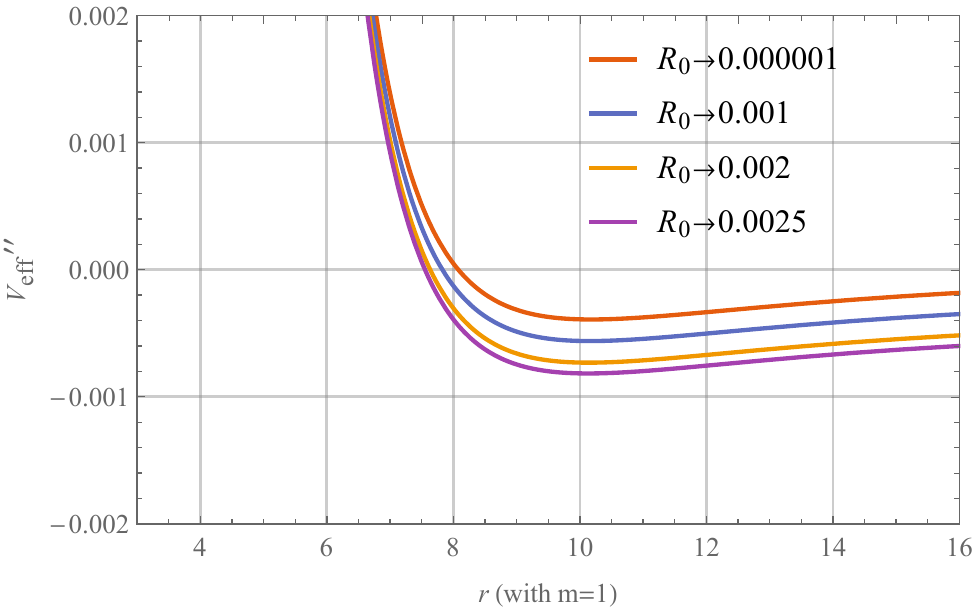}}
    \caption{Plot of (a) $\frac{L^2}{\mu^2}$, \;(b) $V_{\text{eff}}(r)$, \;(c) $V_{\text{eff}}'(r)$, \;(d) $V_{\text{eff}}''(r)$ as a function of $r$ for $R_0=10^{-6}, 0.001, 0.002, 0.0025$ with $m=1$} \label{fig:rpanel_1}
\end{figure}

Through \figref{fig:rpanel_1}, we can illustrate the nature of different properties we computed earlier like, angular momentum \eqref{eq:timelike_angular_momentum}, effective potential \eqref{eq:effective_potential} and its derivaties as a function of our approximated ISCO solution.

\section{Conclusion}
In this article, we have developed a solution for a static spherically symmetric solution under the metric formalism of $\fR$ gravity.
Subsequently, we employed the aforementioned solution and compared it with the Schwarzschild-de Sitter solution in order to determine the values of the constants $C_1$ and $C_2$ as $-2m$ and 0, respectively.
The selection of these particular values was essential for obtaining the solutions for black hole and cosmological horizons for our near-extreme Schwarzschild-dS case.
We confirmed that these solutions converge to Einstein's standard general relativity when the limits of $m\rightarrow1$ and $R_0\rightarrow0$ are applied.
Additonally, using these constants we computed the existence constraints for extreme and hyperextreme Schwarzschild-de Sitter blackholes.

It is apparent to observe that our static spherically symmetric line-element \eqref{eq:sss_metric_horizon} reduces to Schwarzschild's line element for $r\rightarrow0$ or for near-zero scalar curvature.
On the other hand, it becomes the de Sitter line element when $r\rightarrow\infty$ or there's curvature present.

In section 5, we calculated the solutions for circular orbits and demonstrated their stability. Subsequently, we were able to approximate these marginally stable circular orbit as the innermost stable circular orbit.
It is evident from \figref{fig:risco} that the approximated ISCO solution is quadratic with $R_0$ with $m$ being a constant and linear with $m$ with $R_0$ being a constant.
In addition, we were able to compute the nature of angular momentum, effective potential and its derivatives using the approximated ISCO solution in \figref{fig:rpanel_1}.

\nocite{*}
\bibliographystyle{unsrt}
\bibliography{citations}

\end{document}